\documentclass[iop]{emulateapj}

\usepackage{graphicx,natbib,amsmath}

\newcommand{\csb}{\ensuremath{c_{\mathrm{SB}}}}
\newcommand{\chandra}{{\it Chandra}}
\newcommand{\Msun}{\ensuremath{M_\sun}}

\begin{document}

\title{High-Redshift Cool-Core Galaxy Clusters Detected via the Sunyaev--Zel'dovich Effect in the South Pole Telescope Survey}

\altaffiltext{\Munich}{Department of Physics,
Ludwig-Maximilians-Universit\"{a}t,
Scheinerstr.\ 1, D-81679 M\"{u}nchen,
Germany}
\altaffiltext{\UChicago}{University of Chicago,
5640 South Ellis Avenue, Chicago, IL 60637, USA}
\altaffiltext{\CfA}{Harvard-Smithsonian Center for Astrophysics,
60 Garden Street, Cambridge, MA 02138, USA}
\altaffiltext{\MIT}{MIT Kavli Institute for Astrophysics and Space
Research, Massachusetts Institute of Technology, 77 Massachusetts Avenue,
Cambridge, MA 02139, USA}
\altaffiltext{\Harvard}{Department of Physics, Harvard University, 17 Oxford Street, Cambridge, MA 02138, USA}
\altaffiltext{\ExcellenceCluster}{Excellence Cluster Universe,
Boltzmannstr.\ 2, D-85748 Garching, Germany}
\altaffiltext{\KICPChicago}{Kavli Institute for Cosmological Physics,
University of Chicago,
5640 South Ellis Avenue, Chicago, IL 60637, USA}
\altaffiltext{\PhysicsUChicago}{Department of Physics,
University of Chicago,
5640 South Ellis Avenue, Chicago, IL 60637, USA}
\altaffiltext{\Miss}{Department of Physics and Astronomy,
University of Missouri, 
5110 Rockhill Road, Kansas City, MO 64110, USA}
\altaffiltext{\EFIChicago}{Enrico Fermi Institute,
University of Chicago,
5640 South Ellis Avenue, Chicago, IL 60637, USA}
\altaffiltext{\AAUChicago}{Department of Astronomy and Astrophysics,
University of Chicago,
5640 South Ellis Avenue, Chicago, IL 60637, USA}
\altaffiltext{\ANL}{Argonne National Laboratory, 9700 S. Cass Avenue, Argonne, IL, 60439, USA}
\altaffiltext{\NIST}{NIST Quantum Devices Group, 325 Broadway Mailcode 817.03, Boulder, CO 80305, USA}
\altaffiltext{\PUC}{Departamento de Astronomía y Astrofísica, PUC Casilla 306, Santiago 22, Chile}
\altaffiltext{\McGill}{Department of Physics,
McGill University,
3600 Rue University, Montreal, Quebec H3A 2T8, Canada}
\altaffiltext{\Berkeley}{Department of Physics,
University of California, Berkeley, CA 94720, USA}
\altaffiltext{\UFlorida}{Department of Astronomy, University of Florida, Gainesville, FL 32611, USA}
\altaffiltext{\Colorado}{Department of Astrophysical and Planetary Sciences and Department of Physics,
University of Colorado,
Boulder, CO 80309, USA}
\altaffiltext{\NASA}{Department of Space Science, VP62,
NASA Marshall Space Flight Center,
Huntsville, AL 35812, USA}
\altaffiltext{\Davis}{Department of Physics,
University of California, One Shields Avenue, Davis, CA 95616, USA}
\altaffiltext{\LBNL}{Physics Division,
Lawrence Berkeley National Laboratory,
Berkeley, CA 94720, USA}
\altaffiltext{\Caltech}{California Institute of Technology, 1200 E. California Blvd., Pasadena, CA 91125, USA}
\altaffiltext{\Arizona}{Steward Observatory, University of Arizona, 933 North Cherry Avenue, Tucson, AZ 85721, USA}
\altaffiltext{\Michigan}{Department of Physics, University of Michigan, 450 Church Street, Ann
Arbor, MI 48109, USA}
\altaffiltext{\MPE}{Max-Planck-Institut f\"{u}r extraterrestrische Physik,
Giessenbachstr.\ D-85748 Garching, Germany}
\altaffiltext{\CaseWestern}{Physics Department, Center for Education and Research in Cosmology
and Astrophysics,
Case Western Reserve University,
Cleveland, OH 44106, USA}
\altaffiltext{\Minnesota}{Physics Department, University of Minnesota, 116 Church Street S.E., Minneapolis, MN 55455, USA}
\altaffiltext{\STScI}{Space Telescope Science Institute, 3700 San Martin
Dr., Baltimore, MD 21218, USA}
\altaffiltext{\SAIC}{Liberal Arts Department,
School of the Art Institute of Chicago,
112 S Michigan Ave, Chicago, IL 60603, USA}
\altaffiltext{\Yale}{Department of Physics, Yale University, P.O. Box 208210, New Haven,
CT 06520-8120, USA}
\altaffiltext{\BCCP}{Berkeley Center for Cosmological Physics,
Department of Physics, University of California, and Lawrence Berkeley
National Labs, Berkeley, CA 94720, USA}

\def\Munich{1}
\def\UChicago{2}
\def\CfA{3}
\def\MIT{4}
\def\Harvard{5}
\def\ExcellenceCluster{6}
\def\KICPChicago{7}
\def\PhysicsUChicago{8}
\def\Miss{9}
\def\EFIChicago{10}
\def\AAUChicago{11}
\def\ANL{12}
\def\NIST{13}
\def\PUC{14}
\def\McGill{15}
\def\Berkeley{16}
\def\UFlorida{17}
\def\Colorado{18}
\def\NASA{19}
\def\Davis{20}
\def\LBNL{21}
\def\Caltech{22}
\def\Arizona{23}
\def\Michigan{24}
\def\MPE{25}
\def\CaseWestern{26}
\def\Minnesota{27}
\def\STScI{28}
\def\SAIC{29}
\def\Yale{30}
\def\BCCP{31}

\author{
D.~R.~Semler\altaffilmark{\Munich},
R.~\v{S}uhada\altaffilmark{\Munich},
K.~A.~Aird\altaffilmark{\UChicago},
M.~L.~N.~Ashby\altaffilmark{\CfA},
M.~Bautz\altaffilmark{\MIT},
M.~Bayliss\altaffilmark{\CfA,\Harvard},
G.~Bazin\altaffilmark{\Munich,\ExcellenceCluster},
S.~Bocquet\altaffilmark{\Munich},
B.~A.~Benson\altaffilmark{\KICPChicago,\EFIChicago},
L.~E.~Bleem\altaffilmark{\KICPChicago,\PhysicsUChicago},
M.~Brodwin\altaffilmark{\Miss},
J.~E.~Carlstrom\altaffilmark{\KICPChicago,\PhysicsUChicago,\EFIChicago,\AAUChicago,\ANL},
C.~L.~Chang\altaffilmark{\KICPChicago,\EFIChicago,\ANL},
H.~M. Cho\altaffilmark{\NIST},
A.~Clocchiatti\altaffilmark{\PUC},
T.~M.~Crawford\altaffilmark{\KICPChicago,\AAUChicago},
A.~T.~Crites\altaffilmark{\KICPChicago,\AAUChicago},
T.~de~Haan\altaffilmark{\McGill},
S.~Desai\altaffilmark{\Munich,\ExcellenceCluster},
M.~A.~Dobbs\altaffilmark{\McGill},
J.~P.~Dudley\altaffilmark{\McGill},
R.~J.~Foley\altaffilmark{\CfA},
E.~M.~George\altaffilmark{\Berkeley},
M.~D.~Gladders\altaffilmark{\KICPChicago,\AAUChicago},
A.~H.~Gonzalez\altaffilmark{\UFlorida},
N.~W.~Halverson\altaffilmark{\Colorado},
N.~L.~Harrington\altaffilmark{\Berkeley},
F.~W.~High\altaffilmark{\KICPChicago,\AAUChicago},
G.~P.~Holder\altaffilmark{\McGill},
W.~L.~Holzapfel\altaffilmark{\Berkeley},
S.~Hoover\altaffilmark{\KICPChicago,\EFIChicago},
J.~D.~Hrubes\altaffilmark{\UChicago},
C.~Jones\altaffilmark{\CfA},
M.~Joy\altaffilmark{\NASA},
R.~Keisler\altaffilmark{\KICPChicago,\PhysicsUChicago},
L.~Knox\altaffilmark{\Davis},
A.~T.~Lee\altaffilmark{\Berkeley,\LBNL},
E.~M.~Leitch\altaffilmark{\KICPChicago,\AAUChicago},
J.~Liu\altaffilmark{\Munich,\ExcellenceCluster},
M.~Lueker\altaffilmark{\Berkeley,\Caltech},
D.~Luong-Van\altaffilmark{\UChicago},
A.~Mantz\altaffilmark{\KICPChicago,\AAUChicago},
D.~P.~Marrone\altaffilmark{\Arizona},
M.~McDonald\altaffilmark{\MIT},
J.~J.~McMahon\altaffilmark{\KICPChicago,\EFIChicago,\Michigan},
J.~Mehl\altaffilmark{\KICPChicago,\AAUChicago},
S.~S.~Meyer\altaffilmark{\KICPChicago,\PhysicsUChicago,\EFIChicago,\AAUChicago},
L.~Mocanu\altaffilmark{\KICPChicago,\AAUChicago},
J.~J.~Mohr\altaffilmark{\Munich,\ExcellenceCluster,\MPE},
T.~E.~Montroy\altaffilmark{\CaseWestern},
S.~S.~Murray\altaffilmark{\CfA},
T.~Natoli\altaffilmark{\KICPChicago,\PhysicsUChicago},
S.~Padin\altaffilmark{\KICPChicago,\AAUChicago,\Caltech},
T.~Plagge\altaffilmark{\KICPChicago,\AAUChicago},
C.~Pryke\altaffilmark{\Minnesota},
C.~L.~Reichardt\altaffilmark{\Berkeley},
A.~Rest\altaffilmark{\STScI},
J.~Ruel\altaffilmark{\Harvard},
J.~E.~Ruhl\altaffilmark{\CaseWestern},
B.~R.~Saliwanchik\altaffilmark{\CaseWestern},
A.~Saro\altaffilmark{\Munich},
J.~T.~Sayre\altaffilmark{\CaseWestern},
K.~K.~Schaffer\altaffilmark{\KICPChicago,\EFIChicago,\SAIC},
L.~Shaw\altaffilmark{\McGill,\Yale},
E.~Shirokoff\altaffilmark{\Berkeley,\Caltech},
J.~Song\altaffilmark{\Michigan},
H.~G.~Spieler\altaffilmark{\LBNL},
B.~Stalder\altaffilmark{\CfA},
Z.~Staniszewski\altaffilmark{\CaseWestern},
A.~A.~Stark\altaffilmark{\CfA},
K.~Story\altaffilmark{\KICPChicago,\PhysicsUChicago},
C.~W.~Stubbs\altaffilmark{\CfA,\Harvard},
A.~van~Engelen\altaffilmark{\McGill},
K.~Vanderlinde\altaffilmark{\McGill},
J.~D.~Vieira\altaffilmark{\KICPChicago,\PhysicsUChicago,\Caltech},
A. Vikhlinin\altaffilmark{\CfA},
R.~Williamson\altaffilmark{\KICPChicago,\AAUChicago},
O.~Zahn\altaffilmark{\Berkeley,\BCCP},
and
A.~Zenteno\altaffilmark{\Munich,\ExcellenceCluster}
}

\begin{abstract}

We report the first investigation of cool-core properties of galaxy clusters
selected via their Sunyaev--Zel'dovich (SZ) effect.  We use 13 galaxy clusters
uniformly selected from 178 deg$^2$ observed with the South Pole Telescope
(SPT) and followed up by the {\it Chandra X-ray Observatory}.  They form an
approximately mass-limited sample ($> 3\times 10^{14} \Msun h^{-1}_{70}$)
spanning redshifts $0.3 < z < 1.1$.  Using previously published X-ray-selected
cluster samples, we compare two proxies of cool-core strength: surface
brightness concentration (\csb) and cuspiness ($\alpha$).  We find that \csb\
is better constrained.  We measure \csb\ for the SPT sample and find several
new $z > 0.5$ cool-core clusters, including two strong cool cores.  This rules
out the hypothesis that there are no $z > 0.5$ clusters that qualify as strong
cool cores at the 5.4$\sigma$ level.  The fraction of strong cool-core clusters
in the SPT sample in this redshift regime is between 7\% and 56\% (95\%
confidence).  Although the SPT selection function is significantly different
from the X-ray samples, the high-$z$ \csb\ distribution for the SPT sample is
statistically consistent with that of X-ray-selected samples at both low and
high redshifts.  The cool-core strength is inversely correlated with the offset
between the brightest cluster galaxy and the X-ray centroid, providing evidence
that the dynamical state affects the cool-core strength of the cluster.  Larger
SZ-selected samples will be crucial in understanding the evolution of cluster
cool cores over cosmic time.

\end{abstract}


\section{Introduction}

Galaxy clusters grow over cosmic time through mergers with other galaxy
clusters as well as through the accretion of gas and individual galaxies from
the surrounding environment.  On timescales of a few Gyr, radiative cooling due
to X-ray emission from the intracluster medium (ICM) would give rise to a
``cooling flow'' to the cluster core \citep{fabian77,mathews78}, if it were not
countered by a heating mechanism.  These cooling flows are not observed;
instead, the cores of some clusters are found to undergo only moderate cooling
\citep{kaastra01,peterson01,tamura01}.  Such galaxy clusters are known as
``cool-core'' clusters \citep{molendi01}.  Other clusters exhibit little to no
cooling in their core (i.e., noncool-core clusters).  These cooling properties
suggest that there must be processes in every cluster that are strong enough to
either regulate cooling flows or completely prevent them.  Such processes are
not fully understood and it is still uncertain how they evolve and affect
cluster formation over time.

The important astrophysical processes that counteract cool-core formation are
typically thought to fall under three broad categories: feedback from active
galactic nuclei (AGNs), preheating of the cluster gas, and cluster mergers.  AGN
feedback in the cluster's brightest cluster galaxy (BCG) has been shown to be
capable of regulating cooling flows in cool-core clusters \citep[see][for
reviews]{fabian12,mcnamara12}.  In some cases, the feedback may be strong
enough to disrupt a cool core completely \citep[e.g.,][]{mcdonald11b}, though
this phenomenon is likely limited to lower mass galaxy clusters and groups.
Additionally, AGN feedback may drive turbulence in the ICM.  This has been
shown to suppress heat-flux-driven buoyancy instabilities, resulting in
effective transfer of heat from the outer radii and disrupting the cool core
\citep{parrish12b}.  Heating of the intracluster gas during early stages of the
cluster has been shown to affect the formation of cool cores as well
\citep[e.g.,][]{kaiser91,mccarthy08,sun09b}.  Cluster mergers can also disrupt
cool cores by shock-heating and turbulent mixing \citep{leccardi10,rossetti11},
a process that has been reproduced in simulations
\citep[e.g][]{mcglynn84,gomez02b,zuhone10}.  Whether a merger can destroy a
cool core likely depends on the strength of the cool core, the mass ratio of
the merging clusters, and the geometry of the impact.

Studying the evolution of clusters can provide insight regarding the relative
importance of these processes in cool-core formation.  Given that cool cores
develop over a central cooling time of typically a few Gyr, one expects there
to be fewer cool cores at times closer to the epoch of galaxy cluster
formation.  Simulations predict a significantly higher cluster merger rate in
the past \citep{gottloeber01}.  If mergers play a strong role in the disruption
of cool-core galaxy clusters, this too suggests that the fraction of galaxy
clusters with cool cores should be lower at high redshifts than in local
samples.  Indeed, studies of the evolution of the cool-core fraction find a
much lower fraction at $z = 0.5$ \citep{mcdonald11} and a significant dearth in
{\it strong} cool cores at $z > 0.5$ \citep{vikhlinin07,santos10,samuele11}.
To date, only a small number of $z > 0.5$ galaxy clusters with possible strong
cool cores have been reported \citep[e.g.,][]{siemiginowska10,russell12}, with
the most dramatic, confirmed strong cool core coming from the South Pole
Telescope (SPT) survey \citep{mcdonald12}.

Understanding cool-core evolution is complicated by selection biases of cluster
samples.  One can generically expect an X-ray-selected sample to be biased
toward selecting cool-core galaxy clusters \citep{hudson10,eckert11} due to
their higher X-ray surface brightness and luminosity as compared to a
noncool-core galaxy cluster of the same mass \citep[e.g.,][]{ohara06}.  However,
there are competing effects due to X-ray emission from AGNs, which are expected
to be more prevalent at higher redshifts \citep{russell12}.  The bias may be
complicated further by the ways in which different cluster-finding algorithms
differentiate between point sources (e.g., AGNs, X-ray binaries) and extended
sources (e.g., nearby galaxies, groups, and clusters).  Furthermore, the
classification of the cool-core strength of a cluster varies between surveys of
different angular resolution and the method used to characterize the cool core.

Given the complex effects associated with X-ray selection, it is important to
investigate the cool-core fraction and its evolution using an independent
selection method.  In this paper, we study the cool-core properties of galaxy
clusters selected from their Sunyaev-Zel'dovich (SZ) effect \citep{sunyaev72}
signature in the SPT survey.  At $z > 0.3$, this SPT selection is nearly
redshift-independent and nearly constant in mass \citep[e.g.,][]{reichardt12c}.
The SZ effect selection is expected to be relatively insensitive to
non-gravitational physics \citep{nagai06}, the dynamical state of clusters
\citep{jeltema08}, radio contamination from point sources and BCGs
\citep{lin09}, and the presence of cool cores \citep{Motl05}.  Therefore, an
SZ-cluster survey is expected to be a useful tool to study the redshift
evolution of galaxy cluster properties.  This work provides the first glimpse
of the cool-core properties of a sample of galaxy clusters selected from the SZ
effect.

The layout of this paper is as follows.  In Section~\ref{sec:observations}, we
detail the observations used and describe the data reduction procedures.  In
Section~\ref{sec:character-cc}, we present the steps used to make our
measurements as well as compare two methods used to characterize cool-core
strengths that are suitable for high-redshift clusters.  In
Section~\ref{sec:cc-evolution}, we present the results of our measurements and
investigate the implications for the cool-core fraction at high redshifts.  In
Section~\ref{sec:bcg-offset}, we investigate the relationship between a
cluster's cool-core strength and the offset of its BCG.  Finally, in
Section~\ref{sec:conclusions}, we conclude our analyses and present future
studies and applications.

In this analysis, we assume the best-fit WMAP7+BAO+$H_0$ flat $\Lambda$CDM
cosmology \citep{komatsu11} with Hubble parameter $H_0 =
70.4$~km~s$^{-1}$~Mpc$^{-1}$, matter density $\Omega_M = 0.272$, and dark
energy density $\Omega_\Lambda = 0.728$.

\section{Observations}\label{sec:observations}

We analyze \chandra\ data for two different cluster samples, an SZ-selected
sample from \citet[][hereafter A11]{andersson11} and an X-ray-selected
sample from \citet[][hereafter V09]{vikhlinin09b}.  In this section, we
describe each cluster sample and the \chandra\ X-ray data reduction.

\subsection{Cluster Samples}\label{sec:cluster-samples}

We first describe the SZ-selected sample from A11.  This cluster sample is a
subset of the SPT cluster catalog described by \citet{vanderlinde10}, which
consists of 21 clusters selected by their SZ-significance ($\xi$) from
$178~\mathrm{deg}^2$ of sky surveyed by the SPT.  From the subsample of 17
clusters at $z > 0.3$ and $\xi > 5.45$, 15 clusters were selected for X-ray
observations, the results of which are discussed in A11 and \citet{benson11}.
Of the 15 observations, 13 were carried out by \chandra\ and two were carried
out by {\it XMM-Newton}.  For this study, we use the 13 clusters with \chandra\
observations because only \chandra\ provides the spatial resolution needed for
our analysis.  These 13 clusters form a {\it nearly} complete mass-limited
sample (called the ``SPT sample'').

In addition to these, we analyze SPT-CL J2106$-$5844, a massive galaxy cluster
at $z = 1.13$ discovered by the SPT survey \citep{foley11}.  However, this
cluster is not included in analyses involving the distribution of cool-core
strengths (Section~\ref{sec:csb-dist}) as it is not a member of the
mass-limited data set.

We also analyze 41 galaxy clusters, based largely from the high-redshift sample
of the Chandra Cluster Cosmology,
Project\footnote{http://hea-www.harvard.edu/400d/cosm/} known hereafter as the
CCCP high-$z$ sample (V09).  This X-ray-selected sample is the subset of
clusters in the 400d survey \citep{burenin07} at $z \geq 0.35$ and above a
redshift-dependent X-ray-luminosity threshold.  We include five additional
clusters from the 400d survey with \chandra\ data available not presented in
V09.\footnote{These clusters are included to increase the sample size as well
as to remain consistent with the data set presented in \citet{santos10}.}

\subsection{\chandra\ Data Reduction}\label{sec:chandra-reductions}

The \chandra\ observations used in this study are listed in Table
\ref{tab:obs}.  Data are reduced using the \chandra\ software version
\texttt{CIAO 4.4} and \texttt{CALDB 4.4.8}.  For all observations, Level $=2$
event files are generated with the \texttt{chandra\_repro} script.  Exposure
corrections are applied using \texttt{fluximage} in the 0.5--2~keV band with
exposure maps calculated at 1.5~keV.  We employ a two-step procedure in
removing point sources.  First, candidate point sources are identified using
\texttt{wavdetect}.  The results are then visually inspected and false
detections within the cluster are ignored.  Using \texttt{dmfilth}, proper
detections are replaced with a level determined from an elliptical annulus
centered on the point source.  Background levels are determined in each
observation from several regions located on the same chip as the cluster source
without point source detections.  The regions are positioned far enough from
the cluster emission so as to contain negligible cluster photons and are large
enough to adequately sample the background level.

The results enable measurements to be made on the background-subtracted,
exposure-corrected, flux images in the 0.5-2~keV band where each pixel
is in units of photons~cm$^{-2}$~s$^{-1}$.  Count-rate errors are determined
based on Poisson statistics and propagated in the standard way.

\begin{deluxetable}{lc}
\tabletypesize{\scriptsize}
\tablecaption{\chandra\ Observation IDs}
\tablewidth{0pt}
\tablehead{
    \colhead{Cluster} & \colhead{Observation ID} 
    }
\startdata
SPT-CL J0000$-$5748 & 9335 \\
SPT-CL J0509$-$5342 & 9432 \\
SPT-CL J0516$-$5430 & 9331 \\
SPT-CL J0528$-$5300 & 12092, 10862, 11996, 9341, 11874 \\
SPT-CL J0533$-$5005 & 11748, 12001 \\
SPT-CL J0546$-$5345 & 9336, 9332, 10864, 10851 \\
SPT-CL J0551$-$5709 & 11871 \\
SPT-CL J2106$-$5844 & 12180 \\
SPT-CL J2331$-$5051 & 11738, 9333 \\
SPT-CL J2337$-$5942 & 11859 \\
SPT-CL J2341$-$5119 & 11799, 9345 \\
SPT-CL J2342$-$5411 & 11870, 11741, 12014 \\
SPT-CL J2355$-$5056 & 11998, 11746 \\
SPT-CL J2359$-$5009 & 9334, 11742, 11864 \\
\\[-2ex]
\tableline
\\[-1ex]
CL J0340$-$2823 & 5780 \\
CL J0302$-$0423 & 5782 \\
CL J1212+2733 & 5767 \\
CL J0350$-$3801 & 7227 \\
CL J0318$-$0302 & 5775 \\
CL J1514+3636 & 800 \\
CL J0159+0030 & 5777 \\
CL J0958+4702 & 5779 \\
CL J0809+2811 & 5774 \\
CL J1416+4446 & 541 \\
CL J1312+3900 & 5781 \\
CL J1003+3253 & 5776 \\
CL J0141$-$3034 & 5778 \\
CL J1701+6414 & 547 \\
CL J1641+4001 & 3575 \\
CL J0522$-$3624 & 4926, 5837 \\
CL J1222+2709 & 5766 \\
CL J0355$-$3741 & 5761 \\
CL J0853+5759 & 4925, 5765 \\
CL J0333$-$2456 & 5764 \\
CL J0926+1242 & 4929, 5838 \\
CL J0030+2618 & 5762 \\
CL J1002+6858 & 5773 \\
CL J1524+0957 & 1664 \\
CL J1357+6232 & 5763, 7267 \\
CL J1354$-$0221 & 4932, 5835 \\
CL J1117+1744 & 4933, 5836 \\
CL J1120+2326 & 3235 \\
CL J0216$-$1747 & 5760, 6393 \\
CL J0521$-$2530 & 5758 \\
CL J0956+4107 & 5294, 5759 \\
CL J0328$-$2140 & 5755, 6258 \\
CL J1120+4318 & 5771 \\
CL J1334+5031 & 5772 \\
CL J0542$-$4100 & 914 \\
CL J1202+5751 & 5757 \\
CL J0405$-$4100 & 7191 \\
CL J1221+4918 & 1662 \\
CL J0230+1836 & 5754 \\
CL J0152$-$1358 & 913 \\
CL J1226+3332 & 3180, 5014
\enddata
\label{tab:obs}
\end{deluxetable}

\section{Measuring Cool Cores with Imaging Data}\label{sec:character-cc}

\citet{hudson10} compare several X-ray estimators of cool-core strength that
are applied to a cluster sample with a range of X-ray data quality and
redshifts.  They find that the X-ray cuspiness \citep[$\alpha$;][]{vikhlinin07}
and the surface brightness concentration \citep[\csb;][]{santos08} are
promising cool-core estimators for high-redshift clusters with observations
containing relatively few X-ray photons.  In the low-redshift regime, they find
the central cooling time, $t_\mathrm{cool}$ to be the best cool-core estimator
based on the strength of its bimodality.  Both \csb\ and $\alpha$ have been
shown to correlate well with the central cooling time
\citep{santos08,hudson10}.  In this section, we compare the $\alpha$ and \csb\
parameters to the central cooling time for a wider sample of clusters.  We also
calculate \csb\ for the portion of the CCCP high-$z$ sample not previously
published (i.e., clusters at $z < 0.5$).

\subsection{Calculating $\alpha$ and \csb}\label{sec:param-calc}

The cuspiness is defined as the slope of the gas density $\rho_g$ $$ \alpha
\equiv -\frac{d\log{\rho_g}}{d\log{r}}, $$ where the function is evaluated at
radius $r = 0.04 r_{500}$, $r_{500}$ being the radius at which the mean density
of the enclosed mass is 500 times that of the critical density at the object's
redshift \citep{vikhlinin07}.  This radius is close enough to the cluster core
to sample the areas of strongest cooling, while still being far enough to avoid
any flattening of the density profile caused by feedback from a central AGN.
Calculations of $\alpha$ for the CCCP high-$z$ sample are derived from the
X-ray surface brightness fits and X-ray centers used in V09.  

The \csb\ is defined as the ratio of the soft X-ray flux, $F$, within the inner
40~kpc to the inner 400~kpc $$ \csb \equiv
\frac{F_{r<40~\mathrm{kpc}}}{F_{r<400~\mathrm{kpc}}} $$ \citep{santos08}.
These radii were chosen because they provide the largest separation of \csb\
values between cool-core and noncool-core clusters.  Previous studies have used
either the 0.5--5.0~keV energy band or the 0.5--2.0~keV energy band for
determining \csb.  We follow \citet{santos10} in using the 0.5--2.0~keV band
for our \csb\ measurements.

To calculate \csb\ in the clusters described in Section
\ref{sec:cluster-samples}, we first estimate the centroid of the X-ray emission
for each galaxy cluster.  This is determined by iterating the centroid within
an 80~pixel ($\sim$40$\arcsec$) radius, initially centered on the approximate
location of the centroid.  For each iteration, the centroid is calculated with
the image weighted by $r^{-1/2}$, where $r$ is radial distance to the center of
the previous iteration.  For clusters with multiple observations, an inverse
variance weighted final \csb\ is determined using the \csb\ values measured in
the individual observations.  In these clusters, the X-ray centroid is taken to
be the mean centroid of all observations, without any weighting applied.

When comparing X-ray measurements across different redshift regimes, a
$K$-correction is commonly applied to account for the redshift dependence of
the flux in a given band.  This is generally a small effect for the \csb\
parameter \citep{santos10}.  We assume it's negligible for this work because we
are primarily interested in comparing several galaxy cluster samples at
similarly high redshifts.

Using the above analysis, we report $\alpha$ and \csb\ for the CCCP high-$z$
sample in Table~\ref{tab:cccp}, with \csb\ values for clusters at $z \geq 0.5$
from \citet{santos10}.

\begin{deluxetable}{lllr}
\tabletypesize{\scriptsize}
\tablewidth{0pt}
\tablecaption{CCCP High-$z$ Galaxy Cluster Measurements}
\tablehead{ \colhead{Cluster} & \colhead{$z$} &\colhead{\csb} & \colhead{$\alpha$} \\
& \colhead{(1)} & &
}
\startdata
CL J0340$-$2823\tablenotemark{a} & $0.35$ & $0.114 \pm 0.012$ & $0.920 \pm 0.068$ \\ 
CL J0302$-$0423 & $0.35$ & $0.374 \pm 0.024$ & $1.349 \pm 0.053$ \\
CL J1212+2733 & $0.35$ & $0.036 \pm 0.005$ & $0.389 \pm 0.133$ \\
CL J0350$-$3801 & $0.36$ & $0.073 \pm 0.012$ & $0.126 \pm 0.176$ \\
CL J0318$-$0302 & $0.37$ & $0.039 \pm 0.007$ & $0.078 \pm 0.102$ \\
CL J1514+3636\tablenotemark{a} & $0.37$ & $0.276 \pm 0.014$ & $1.185 \pm 0.049$ \\
CL J0159+0030 & $0.39$ & $0.175 \pm 0.017$ & $1.102 \pm 0.071$ \\
CL J0958+4702 & $0.39$ & $0.144 \pm 0.016$ & $0.751 \pm 0.148$ \\
CL J0809+2811 & $0.40$ & $0.028 \pm 0.006$ & $0.021 \pm 0.089$ \\
CL J1416+4446 & $0.40$ & $0.149 \pm 0.012$ & $1.008 \pm 0.064$ \\
CL J1312+3900 & $0.40$ & $0.046 \pm 0.009$ & $0.026 \pm 0.036$ \\
CL J1003+3253 & $0.42$ & $0.216 \pm 0.022$ & $1.369 \pm 0.188$ \\
CL J0141$-$3034 & $0.44$ & $0.058 \pm 0.014$ & $0.563 \pm 0.284$ \\
CL J1701+6414 & $0.45$ & $0.155 \pm 0.013$ & $1.189 \pm 0.043$ \\
CL J1641+4001 & $0.46$ & $0.087 \pm 0.011$ & $0.695 \pm 0.314$ \\
CL J0522$-$3624 & $0.47$ & $0.048 \pm 0.008$ & $0.671 \pm 0.102$ \\
CL J1222+2709 & $0.47$ & $0.115 \pm 0.014$ & $0.743 \pm 0.132$ \\
CL J0355$-$3741 & $0.47$ & $0.096 \pm 0.013$ & $0.877 \pm 0.076$ \\
CL J0853+5759 & $0.48$ & $0.025 \pm 0.007$ & $0.190 \pm 0.208$ \\
CL J0333$-$2456 & $0.48$ & $0.035 \pm 0.007$ & $0.395 \pm 0.173$ \\
CL J0926+1242 & $0.49$ & $0.092 \pm 0.010$ & $0.744 \pm 0.116$ \\
CL J0030+2618 & $0.50$ & $0.040 \pm 0.011$\tablenotemark{b} & $0.358 \pm 0.155$ \\
CL J1002+6858 & $0.50$ & $0.060 \pm 0.012$\tablenotemark{b} & $0.185 \pm 0.147$ \\
CL J1524+0957 & $0.52$ & $0.032 \pm 0.006$\tablenotemark{b} & $0.056 \pm 0.698$ \\
CL J1357+6232 & $0.53$ & $0.054 \pm 0.010$\tablenotemark{b} & $0.632 \pm 0.113$ \\
CL J1354$-$0221 & $0.55$ & $0.043 \pm 0.009$\tablenotemark{b} & $0.035 \pm 0.434$ \\
CL J1117+1744\tablenotemark{a} & $0.55$ & $0.041 \pm 0.010$\tablenotemark{b} & $0.322 \pm 0.221$ \\
CL J1120+2326 & $0.56$ & $0.027 \pm 0.011$\tablenotemark{b} & $0.280 \pm 0.127$ \\
CL J0216$-$1747\tablenotemark{a} & $0.58$ & $0.055 \pm 0.014$\tablenotemark{b} & $0.428 \pm 0.203$ \\
CL J0521$-$2530\tablenotemark{a} & $0.58$ & $0.046 \pm 0.007$\tablenotemark{b} & $0.500 \pm 0.225$ \\
CL J0956+4107 & $0.59$ & $0.040 \pm 0.007$\tablenotemark{b} & $0.026 \pm 0.007$ \\
CL J0328$-$2140 & $0.59$ & $0.062 \pm 0.009$\tablenotemark{b} & $0.513 \pm 0.185$ \\
CL J1120+4318 & $0.60$ & $0.063 \pm 0.005$\tablenotemark{b} & $0.679 \pm 0.101$ \\
CL J1334+5031 & $0.62$ & $0.068 \pm 0.017$\tablenotemark{b} & $0.381 \pm 0.228$ \\
CL J0542$-$4100 & $0.64$ & $0.043 \pm 0.007$\tablenotemark{b} & $0.454 \pm 0.136$ \\
CL J1202+5751 & $0.68$ & $0.030 \pm 0.008$\tablenotemark{b} & $0.009 \pm 0.121$ \\
CL J0405$-$4100 & $0.69$ & $0.073 \pm 0.009$\tablenotemark{b} & $0.293 \pm 0.111$ \\
CL J1221+4918 & $0.70$ & $0.026 \pm 0.006$\tablenotemark{b} & $0.049 \pm 0.032$ \\
CL J0230+1836 & $0.80$ & $0.036 \pm 0.009$\tablenotemark{b} & $0.161 \pm 1.171$ \\
CL J0152$-$1358 & $0.83$ & $0.027 \pm 0.008$\tablenotemark{b} & $0.102 \pm 0.364$ \\
CL J1226+3332 & $0.89$ & $0.086 \pm 0.007$\tablenotemark{b} & $0.333 \pm 0.075$
\enddata
\tablecomments{(1) Redshifts from \citet{vikhlinin09b}.}
\tablenotetext{a}{Cluster not a member of the original CCCP high-$z$ set presented in V09.}
\tablenotetext{b}{\csb\ values from \citet{santos10}.}
\label{tab:cccp}
\end{deluxetable}

In the following analyses, we adopt the three different cool-core regimes
defined previously for these parameters \citep{vikhlinin07,santos08}: the
noncool-core regime ($\csb < 0.075$; $\alpha < 0.5$), the moderate regime
($0.075 < \csb < 0.155$; $0.5 < \alpha < 0.7$), and the strong cool-core regime
($\csb > 0.155$; $\alpha > 0.7$).

\subsection{Comparison of \csb\ and $\alpha$}\label{sec:comparison}

\begin{figure}
\begin{center}
\includegraphics[width=0.49\textwidth]{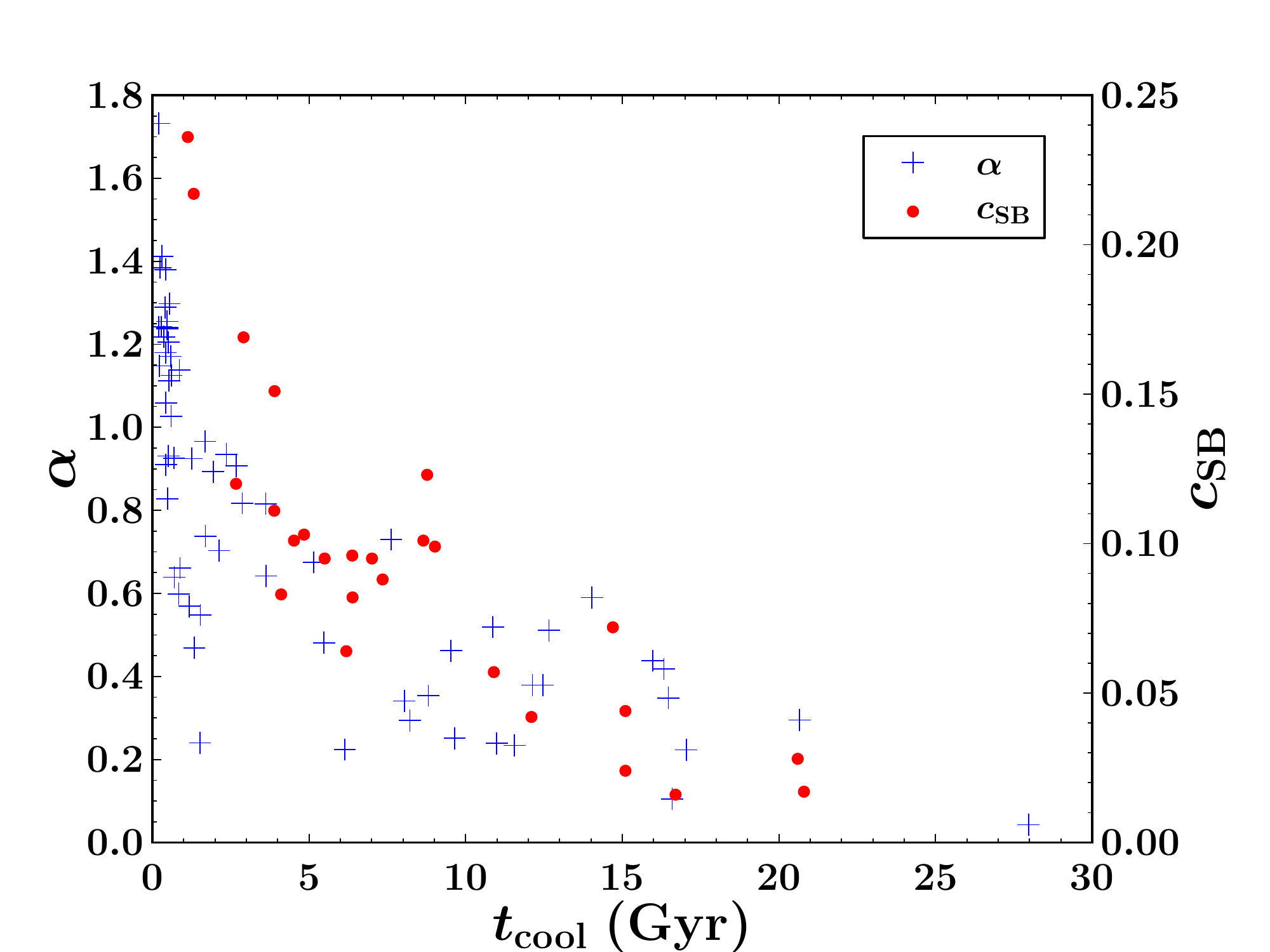}

\caption{Comparison of $\alpha$ and \csb\ with $t_\mathrm{cool}$.  Galaxy
clusters with $\alpha$ and $t_\mathrm{cool}$ values are taken from
\citet{hudson10} and galaxy clusters with \csb\ and $t_\mathrm{cool}$ values
are taken from \citet{santos08}.}

\label{tcool}
\end{center}
\end{figure}

We study the performances of $\alpha$ and \csb\ by first relating the two
parameters to $t_\mathrm{cool}$ in a sample of low-redshift galaxy clusters.
In Figure \ref{tcool}, we compare the measurements of \csb\ and
$t_\mathrm{cool}$ from \citet{santos08} with those of $\alpha$ and
$t_\mathrm{cool}$ from \citet{hudson10}.  We note that \citet{santos08}
calculate \csb\ in a different band of 0.5--5.0~keV, however this should not
qualitatively change our conclusions because the bulk of ICM emission is in the
0.5--2.0~keV band.  We find $\alpha$ and $t_\mathrm{cool}$ to have a
Spearman's rank correlation coefficient of $\rho_{\alpha} = -0.88$.  The
associated $p$-value corresponds to a probability of $4.4 \times 10^{-22}$
that there is no correlation between the two parameters.  The results are
similar for \csb\ and $t_\mathrm{cool}$; their Spearman's rank correlation
coefficient is $\rho_{\csb} = -0.84$ with a $p$-value of $6.3 \times 10^{-8}$.

\begin{figure}
\begin{center}
\includegraphics[width=0.49\textwidth]{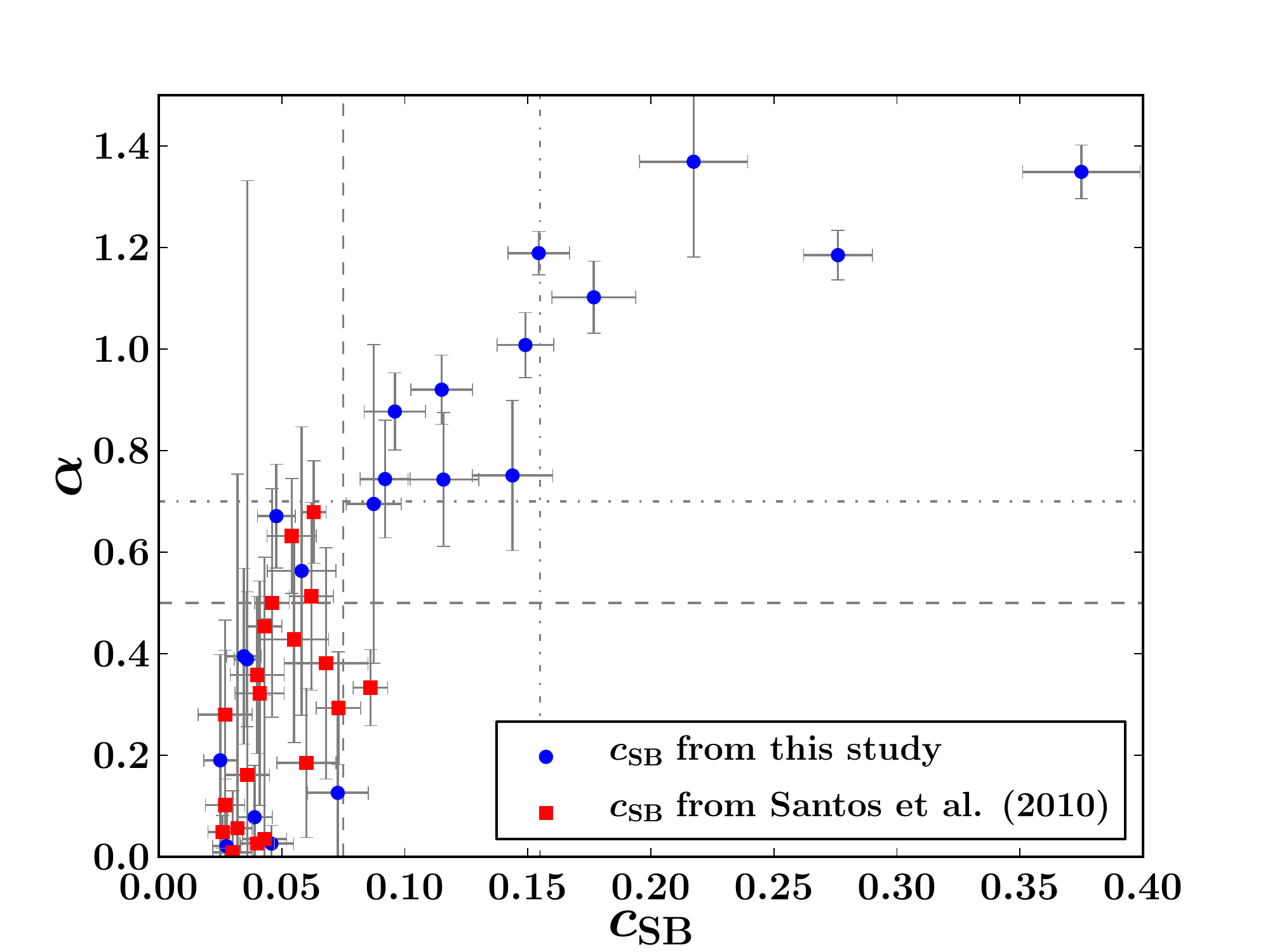}

\caption{Correlation between $\alpha$ and \csb\ for the CCCP high-$z$
clusters.  Values for $\alpha$ are from \citet{vikhlinin07}.  Values for \csb\
with $z < 0.5$ are reported in this study.  Values for \csb\ with $z \geq 0.5$
are from \citet{santos10}.  The dashed lines correspond to the boundaries
between noncool cores and moderate cool cores.  The dash-dotted lines
correspond to the boundaries between moderate cool cores and strong
cool cores.}

\label{alpha_c_SB}
\end{center}
\end{figure}

We compare the two parameters directly for the galaxy clusters in the CCCP
high-$z$ set.  As shown in Figure~\ref{alpha_c_SB}, the $\alpha$ parameter
exhibits larger fractional errors than \csb\ in the CCCP high-$z$ sample.  This
is particularly prevalent for low $\alpha$, indicating that the parameter is
less constrained in the noncool-core regime.  Even excluding the 6 clusters
where the signal-to-noise ratio (S/N) is less than 0.5, the median S/N in
$\alpha$ is 2.9.  However, for all clusters in this sample, the median S/N in
\csb\ is 7.8.

As shown in Figure~\ref{alpha_c_SB}, the two parameters do not always provide
the same classification for a cluster's cool-core strength.  The classification
thresholds for \csb\ and $\alpha$ were defined in \citet{santos08} and
\citet{vikhlinin07}, respectively.  The definitions were based on separate
data sets using characteristics of each parameters' distribution.  Therefore,
one would not necessarily expect perfect agreement between the two parameters.
Regardless, using these classifications, all strong cool-core clusters
identified by \csb\ are also classified as such when using $\alpha$.
Therefore, the \csb\ parameter provides a more conservative threshold for
identifying strong cool-core clusters.

Throughout the rest of the paper, we will use the \csb\ parameter to
characterize the cool-core strength of both the X-ray and SZ-selected samples.
The X-ray observations of these samples have comparable S/N to the samples used
in this section.  Therefore, we expect similar performances from $\alpha$ and
\csb\ as was shown here, where the results suggest that \csb\ is a more robust
and comparably accurate cool-core proxy.

\section{Evolution of the Cool-Core Fraction}\label{sec:cc-evolution}

In this section, we compare the \csb\ distribution as a proxy for the cool-core
fraction for various comparable cluster samples.  We first describe the
X-ray-selected cluster samples previously studied and then go on to present
the results for the SPT sample.  We compare the \csb\ distribution of the
X-ray-selected sample with that of the SPT sample in different redshift
regimes.  Finally, we discuss systematics that could affect the SPT sample.  

\subsection{X-ray-Selected Cluster Samples}

Previously, \citet{santos10} compared the surface brightness concentrations for
57 galaxy clusters.  They compared both low- and high-redshift samples as well
as clusters selected by different X-ray selection methods.  All of the clusters
were first detected in {\it ROSAT} PSPC observations and were later followed up
with \chandra.  About half of the clusters (26) are from the $0.05 \leq z \leq
0.22$ portion of the CCCP low-$z$ set (V09), a flux-limited cluster sample with
many of the same members presented in \citet{edge90}.  The other half is
comprised of clusters in the CCCP high-$z$ set (V09) with $z \geq 0.5$ (20
clusters)\footnote{The 20 clusters from the CCCP high-$z$ sample with $z \geq
0.5$ presented in \citet{santos10} include three clusters not in the original
CCCP high-$z$ sample, as their \chandra\ observations revealed they did not
meet the flux criteria for membership.} and the $z \geq 0.6$ clusters from the
RDCS\footnote{ROSAT Deep Cluster Survey} \citep{rosati98} and the
WARPS\footnote{Wide Angle ROSAT Pointed Survey,
http://asd.gsfc.nasa.gov/Donald.Horner/warps/index.html} \citep{horner08}
samples that have \chandra\ observations (15 Clusters).  One of these clusters,
WARP J1415.1+3612, has an updated \csb\ in \citet{santos12}.  Due to the low
number of counts and similar selection properties in the RDCS and WARPS
surveys, \citet{santos10} group them together in their analysis as RDCS+WARPS.
For consistency, we follow the same practice here.  Note that four clusters are
members of both RDCS+WARPS and the CCCP high-$z$ samples.  There is one
additional high-redshift cluster that has a published \csb\ value, XMMU
J1230.3+1339 \citep{fassbender11b}.

\citet{santos10} compare the \csb\ distributions of the CCCP high-$z$ and
RDCS+WARPS samples using a K-S test.  The result yields a probability of only
0.6\% for the null hypothesis that the two distributions are drawn from the
same parent distribution.  \citet{santos10} argue this is due to a bias of the
detection algorithm used in CCCP against compact clusters with a relatively
high mean surface brightness.  This conflicts with \citet{burenin07}, who
tested the CCCP selection algorithm on a morphologically diverse set of cluster
images, reprojected to redshifts between 0.35 and 0.80.  The results showed
comparable selection efficiencies for all of the cluster morphologies,
indicating the selection method does not present a bias with respect to the
cool-core strength of the cluster.

\subsection{The SPT Sample}

The SPT sample is described in Section~\ref{sec:cluster-samples}.  It consists
of 13 clusters with X-ray observations that have been previously described in
A11 and \citet{benson11}.  We also provide a \csb\ measurement for SPT-CL
J2106$-$5844, whose other X-ray properties are discussed in \citet{foley11}.
Their \csb\ values are provided in Table~\ref{tab:data}.

In the SPT sample, we find two high-redshift galaxy clusters with surface
brightness concentrations in the strong cool-core regime.  These are among the
first strong cool-core galaxy clusters detected at redshifts beyond $z = 0.5$.
One of the strong cool-core clusters is SPT-CL J2331$-$5051, which lies at $z =
0.58$ and has a \csb\ of $0.214 \pm 0.016$.  The strongest cool-core cluster is
SPT-CL J0000$-$5748 and has a \csb\ of $0.244 \pm 0.023$.  At a redshift of
0.702, corresponding to 7.4 Gyr after the big bang, this cluster is also the
highest redshift strong cool-core cluster found in this work.  These two
clusters rule out the hypothesis that there are no galaxy clusters at $z > 0.5$
classified by \csb\ as having strong cool cores at the 5.4$\sigma$ level.

The X-ray properties of the galaxy cluster SPT-CL J2106$-$5844 were studied in
detail in \citet{foley11}.  With a mass $M_{200} = (1.27 \pm
0.21)~\times~10^{15}~h^{-1}_{70}~\Msun$ and a redshift $z=1.13$, it
is the most massive cluster known at $z > 1$.  They measure a cluster
temperature of $T_X = 11.0^{+2.6}_{-1.9}$~keV and a central temperature of $T_X
= 6.5^{+1.7}_{-1.1}$~keV within $0.17 r_{500}$.  The temperature decrement in
the core suggests moderate cooling.  We measure the surface brightness
concentration $\csb = 0.026 \pm 0.007$, which lies below the moderate cool-core
threshold.

\begin{deluxetable*}{lrrrrr}
\tabletypesize{\scriptsize}
\tablewidth{0pt}
\tablecaption{SPT Galaxy Cluster Measurements}
\tablehead{ \colhead{Cluster} & \colhead{R.A.} & \colhead{Decl.} & \colhead{\csb} & \colhead{$z$} & \colhead{BCG Offset} \\ 
& \colhead{(1)} & \colhead{(1)} & & \colhead{(2)} & \colhead{(3) (kpc)}
}
\startdata
SPT-CL J0000$-$5748 & $0.250$ & $-57.810$ & $0.244 \pm 0.023$ & $0.70$ & $7.4 \pm 3.5$ \\
SPT-CL J0509$-$5342 & $77.338$ & $-53.703$ & $0.109 \pm 0.010$ & $0.46$ & $24.1 \pm 2.9$ \\
SPT-CL J0516$-$5430 & $79.148$ & $-54.506$ & $0.026 \pm 0.005$ & $0.30$ & $107.3 \pm 2.2$ \\
SPT-CL J0528$-$5300 & $82.020$ & $-52.997$ & $0.061 \pm 0.009$ & $0.77$ & $58.1 \pm 3.7$ \\
SPT-CL J0533$-$5005 & $83.405$ & $-50.098$ & $0.013 \pm 0.007$ & $0.88$ & $414.2 \pm 3.8$ \\
SPT-CL J0546$-$5345 & $86.655$ & $-53.759$ & $0.072 \pm 0.010$ & $1.07$ & $40.8 \pm 4.0$ \\
SPT-CL J0551$-$5709 & $87.893$ & $-57.144$ & $0.034 \pm 0.008$ & $0.42$ & $82.0 \pm 2.7$ \\
SPT-CL J2106$-$5844\tablenotemark{a} & $316.518$ & $-58.742$ & $0.027 \pm 0.007$ & $1.13$ & $24.3 \pm 4.1$ \\
SPT-CL J2331$-$5051 & $352.963$ & $-50.865$ & $0.214 \pm 0.016$ & $0.58$ & $4.5 \pm 3.2$ \\
SPT-CL J2337$-$5942 & $354.353$ & $-59.705$ & $0.033 \pm 0.006$ & $0.78$ & $199.4 \pm 3.7$ \\
SPT-CL J2341$-$5119 & $355.301$ & $-51.329$ & $0.092 \pm 0.009$ & $1.00$ & $3.0 \pm 4.0$ \\
SPT-CL J2342$-$5411 & $355.692$ & $-54.185$ & $0.138 \pm 0.013$ & $1.08$ & $8.6 \pm 4.1$ \\
SPT-CL J2355$-$5056 & $358.948$ & $-50.928$ & $0.113 \pm 0.010$ & $0.32$ & $6.7 \pm 2.3$ \\
SPT-CL J2359$-$5009 & $359.931$ & $-50.170$ & $0.035 \pm 0.007$ & $0.78$ & $85.5 \pm 3.7$
\enddata
\label{tab:data}
\tablecomments{(1) Coordinates determined from X-ray centroid. (2) Redshifts from \citet{song12b}. (3) Projected offset between BCG position and X-ray centroid, with BCG positions taken from \citet{song12b}.  The error in offset corresponds to the resolution limit of \chandra\ at the cluster's redshift.}
\tablenotetext{a}{SPT-CL J2106$-$5844 is not included in the analysis of the \csb\ distribution.}
\end{deluxetable*}

\begin{figure}
\begin{center}
\includegraphics[width=0.49\textwidth]{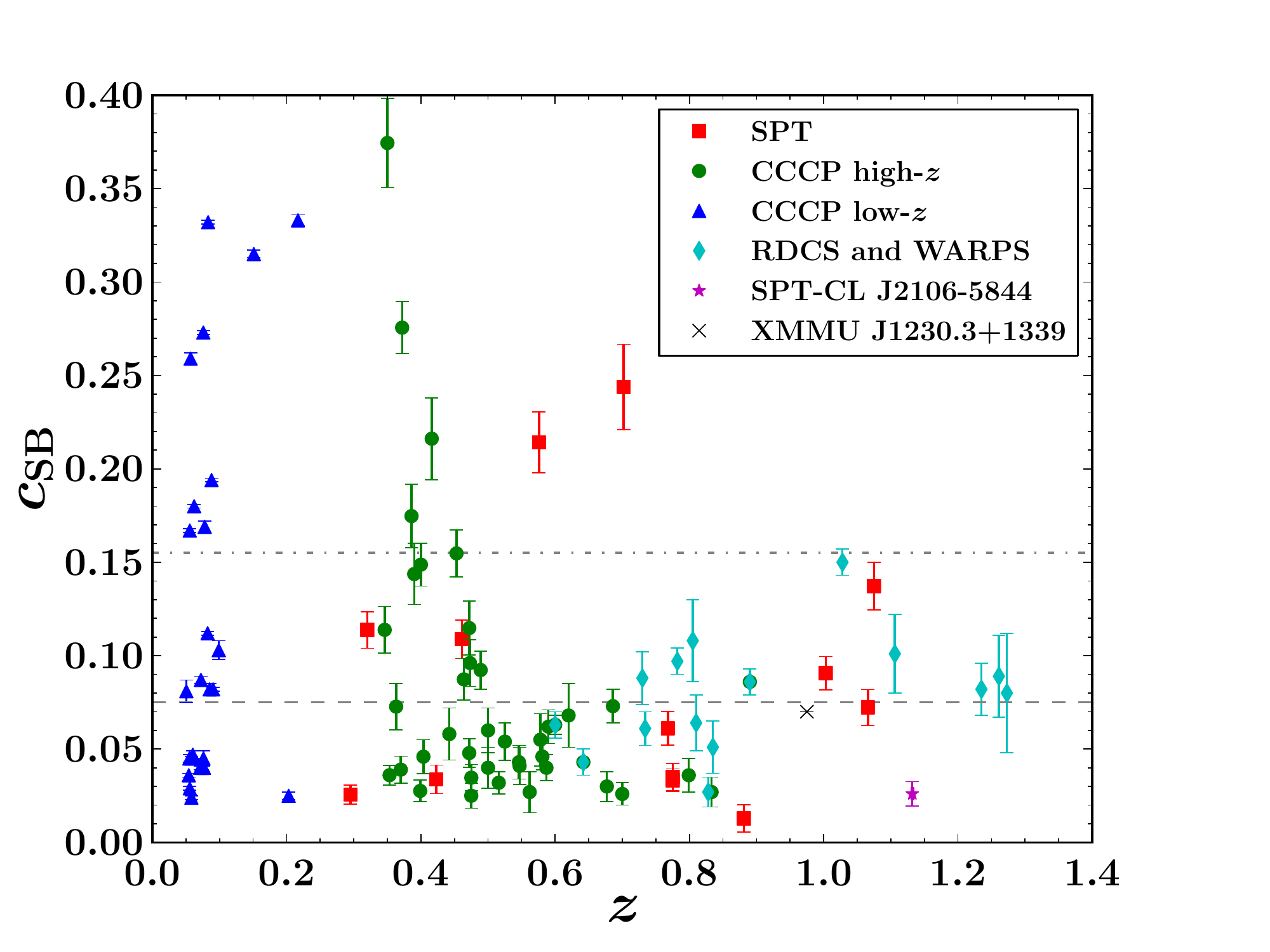}

\caption{Surface brightness concentration as a function of redshift for objects
with a measured \csb\ included in this study.  Four galaxy clusters are members
of two surveys and are shown with symbols from both surveys.  The dashed line
at $\csb = 0.075$ corresponds to the boundary between noncool-core and moderate
cool-core clusters.  The dash-dotted line at $\csb = 0.155$ corresponds to the
boundary between moderate and strong cool-core clusters.}

\label{csbz}
\end{center}
\end{figure}

\subsection{\csb\ Distribution}\label{sec:csb-dist}

In this section, we compare the \csb\ distributions of the SPT and
X-ray-selected high-redshift cluster samples.  To remain consistent with
\citet{vikhlinin07} and \citet{santos10}, we define a redshift break in our
cluster samples at $z = 0.5$, with clusters at $z \geq 0.5$ defined to be
``high-$z$'' and clusters at $z < 0.5$ defined to be ``low-$z$''.  We do not
include SPT-CL J2106$-$5844 in these analyses as it is not part of a uniformly SZ-selected sample with X-ray follow-up, as is approximately the case for the clusters studied in A11.

The distribution of \csb\ as a function of redshift for various X-ray-selected
samples and the SPT sample is in Figure~\ref{csbz}.  At $z > 0.5$, there are 31
clusters in X-ray-selected samples and 9 in the SPT sample.  In this regime,
there are only two strong cool-core clusters, both in the SPT sample.  This
constrains the fraction of strong cool cores at high redshifts in the SPT
sample to between $7\%$ and $56\%$ at a $95\%$ binomial confidence interval
\citep[see][for calculation details]{cameron11}.

\begin{figure}
\begin{center}
\includegraphics[width=0.49\textwidth]{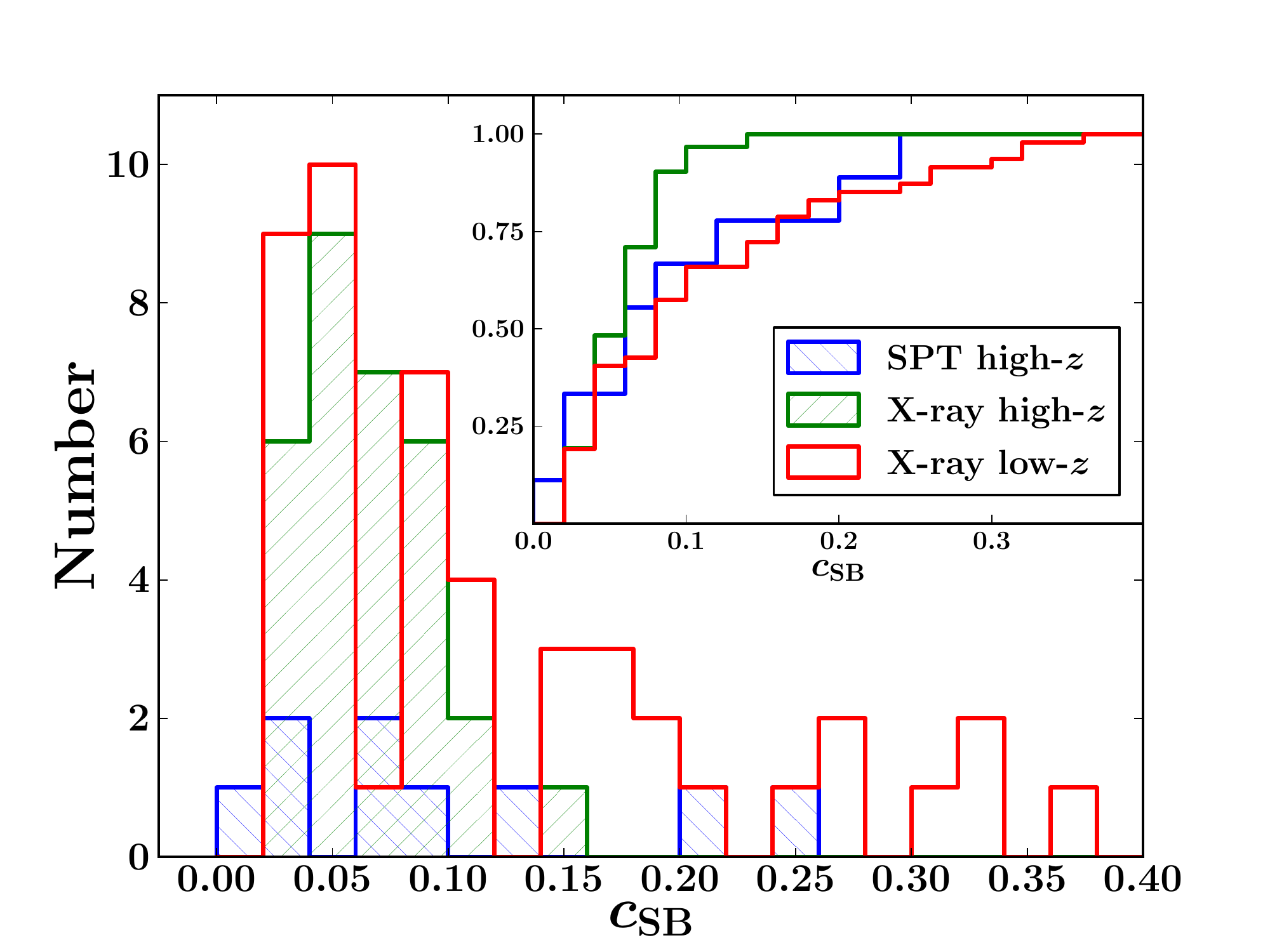}

\caption{Distribution of the surface brightness concentration of the X-ray
high-$z$ sample (the $z \geq 0.5$ clusters in the CCCP high-$z$ sample and the
$z > 0.6$ clusters in the RDCS+WARPS samples; {\it green forward-hatched
areas}), the X-ray low-$z$ sample (the $z < 0.5$ clusters from the CCCP
high-$z$ sample and the CCCP low-$z$ sample; {\it red empty areas}), and the
SPT high-$z$ clusters (the $z > 0.5$ clusters from the SPT sample; {\it blue
back-hatched areas}). {\it Insert}: cumulative distribution of the surface
brightness concentration for the same three samples.}

\label{csb_hist}
\end{center}
\end{figure}

The \csb\ distributions of X-ray and SZ samples, broken up into low-$z$ and
high-$z$ groups, are shown in Figure~\ref{csb_hist}.  We perform a K-S test
between the SPT high-$z$ sample and the X-ray high-$z$ sample.  The result
yields a $p$-value of 0.42, indicating that although the SPT sample contains
the only strong cool-core clusters, the two selection methods show no evidence
for being drawn from different parent distributions.

In order to study the evolution of the cool-core fraction we compare the \csb\
distribution of the SPT high-$z$ sample with that of the X-ray-selected
low-$z$ sample (Figure~\ref{csb_hist})\footnote{We do not study the evolution
solely within the SPT sample because at $z < 0.5$, the sample has only four
clusters.}.  A K-S test between the two data sets results in a $p$-value of
0.87, providing no evidence that these two distributions are drawn from
different parent distributions.  Therefore, although there is a smaller
fraction of strong cool-core clusters in the high-redshift SPT sample, these
results provide no evidence for evolution of the cool-core fraction between the
two redshift regimes.

\subsection{Investigation of AGN Contamination}\label{sec:agn}

Potentially, strong radio emission from AGNs in cool-core clusters could affect
their SZ detections.  Previous investigations suggest that this should not be a
significant concern for samples selected at 150 GHz \citep{lin09,sehgal10}.
A11 investigate the role radio AGNs play specifically for the SPT sample.
Using the Sydney University Molonglo Sky Survey \citep{mauch03} at 843 MHz, A11
identify radio sources in the centers of 7 of the 13 clusters from the SPT
sample.  They extrapolate the radio flux to the SPT bands and find the total
flux of these sources to be negligible compared to the SZ signal.

Measurement of \csb\ can be affected by the presence of an X-ray-loud AGN
within the core radius of the cluster.  Such systems result in inflated \csb\
values due to the AGN's contribution to the core X-ray counts.  In this
section, we investigate AGN contamination in the moderate and strong cool-core
clusters from the SPT sample.

While there are many signatures that a galaxy contains an AGN, there is no
definitive diagnostic that the AGN is emitting X-rays besides direct
observation of X-rays.  Therefore, with the high resolution data provided by
\chandra, the best way of identifying such features is to visually inspect the
images to investigate whether any point sources are within the inner 40~kpc of
the cluster center.  To validate the visual inspection, we also employ a
fully automated source-detection method using the \texttt{CIAO} tool
\texttt{wavdetect}.  We also specifically investigate for the presence of a
point source coincident with the X-ray centroid by comparing the hardness
ratios of the cluster center with an annulus around the center.  The centrally
decreasing temperature of a cool-core cluster means that an uncontaminated core
will have softer emission in the center than in the annulus.  An X-ray AGN,
however, has hard X-ray emission and its presence in the cluster core will
cause the core to have a higher hardness ratio than the annulus.

We utilize \texttt{BEHR} \citep{park06}, a fully Bayesian approach to calculate
hardness ratios that treats photon counts as Poisson statistics with
appropriate error propagation.  We use an X-ray hardness ratio defined as
$(H-S)/(H+S)$ where $H$ corresponds to the X-ray counts in the hard band
(2--8~keV) and $S$ is the X-ray counts in the soft band (0.5--2~keV).  The
hardness ratio of the core is taken in a 2 arcsec aperture centered on the
X-ray centroid.  This aperture size is used as it would capture approximately
90\% of the X-rays from a point source located in its center.  The annulus has
an inner radius of 2 arcsec and an outer radius of 4 arcsec.

In Table~\ref{tab:agn}, we give the results of our checks for AGN
contamination.  Visual inspection of the \chandra\ images reveals no evidence
of an X-ray point source within 40~kpc of any of the cluster centers.  The
source-detection tool \texttt{wavdetect} only detected a source in the center
of one cluster: SPT-CL J0000$-$5748.  In this case, the ratio of the size of the
detection with the size of \chandra's point-spread function at the source
location is 1.8, indicating the detected source is more extended than a point
source (e.g., an AGN) and the emission comes from the cluster's cool-core ICM.
In addition, the core hardness ratios range from $-0.73^{+0.13}_{-0.23}$ to
$-0.48^{+0.07}_{-0.09}$ and are all lower than the hardness ratios in their
surrounding annulus.  These core hardness ratios are also softer than the
hardness ratios of typical AGNs.  For example, \citet{hickox09} report a
hardness ratio of $-0.37\pm0.06$ for a stack of 95 radio-selected AGNs.
Therefore, we find no evidence for AGN contamination in our \csb\ measurements
for the SPT sample.

\begin{deluxetable}{lcccc}
\tabletypesize{\scriptsize}
\tablecaption{AGN Signatures for Moderate and Strong Cool-Core Clusters}
\tablewidth{0pt}
\tablehead{
    \colhead{Cluster} & \colhead{HR$_\mathrm{core}$} & \colhead{HR$_\mathrm{annulus}$} & \multicolumn{2}{c}{Central Source} \\
    & & & \colhead{Wav} & \colhead{Vis} \\
    & \colhead{(1)} & \colhead{(1)} & \colhead{(2)} & \colhead{(3)}
    }
\startdata
SPT-CL J0000$-$5748 & $-0.48^{+0.07}_{-0.09}$ & $-0.39^{+0.07}_{-0.08}$ & Yes & No \\
SPT-CL J0509$-$5342 & $-0.51^{+0.15}_{-0.15}$ & $-0.08^{+0.12}_{-0.12}$ & No & No \\
SPT-CL J2331$-$5051 & $-0.58^{+0.11}_{-0.15}$ & $-0.39^{+0.12}_{-0.14}$ & No & No \\
SPT-CL J2341$-$5119 & $-0.64^{+0.16}_{-0.19}$ & $-0.35^{+0.12}_{-0.13}$ & No & No \\
SPT-CL J2342$-$5411 & $-0.66^{+0.13}_{-0.19}$ & $-0.53^{+0.14}_{-0.18}$ & No & No \\
SPT-CL J2355$-$5056 & $-0.73^{+0.13}_{-0.23}$ & $-0.52^{+0.18}_{-0.18}$ & No & No
\enddata
\tablecomments{(1) Hardness ratio calculated in the cluster core and its surrounding annulus.  A core with a lower hardness ratio than its annulus is consistent with no contamination from an X-ray AGN.  (2) Indication of a source detection by \texttt{wavdetect} at the cluster center.  (3) Indication of an X-ray point source within the inner 40~kpc of the cluster center based on visual inspection of the observations.}
\label{tab:agn}
\end{deluxetable}

\section{Dynamical State and Cool-Core Strengths}\label{sec:bcg-offset}

In this section, we examine the relationship between a cluster's dynamical
state and its cool-core strength.  This can be used to investigate the effect
of mergers on the cool-core properties of our sample.  For this analysis, we
characterize a cluster's dynamical state based on the offset of its BCG from
the center of its X-ray emission \citep{katayama03}.  Having a complete sample
is less important here because the BCG position is not involved in our
selection techniques.  Therefore, in this section, we include SPT-CL
J2106$-$5844 in the SPT sample.

The BCG offsets are determined from the projected distance between the X-ray
centroid (as described for the \csb\ calculation in
Section~\ref{sec:param-calc}) and the BCG positions reported in \citet{song12b}.
Uncertainties in the offsets are given as the resolution for the \chandra\
observations at each cluster's redshift.

\begin{figure}
\begin{center}
\includegraphics[width=0.49\textwidth]{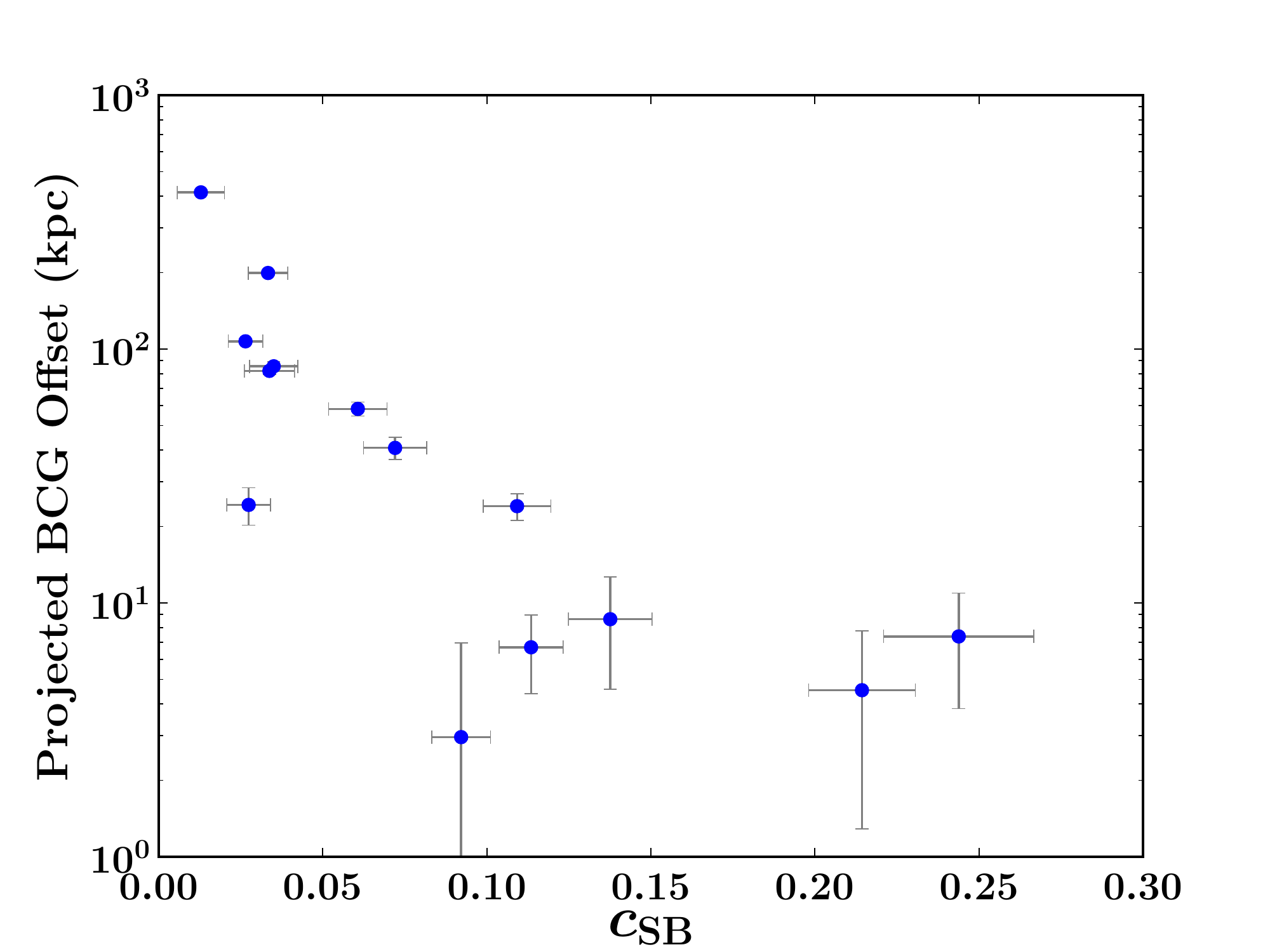}

\caption{Projected offset between the X-ray centroid and the BCG as a function
of \csb.  Errors in the BCG offset correspond to the resolution limit of
\chandra\ at the cluster's redshift.  Data from the SPT sample and SPT-CL
J2106$-$5844.}

\label{offset}
\end{center}
\end{figure}

Although the majority of local clusters exhibit relatively small BCG offsets
\citep{lin04b}, \citet{sanderson09} show that the magnitude of this offset is
anticorrelated with the cool-core strength of the cluster.  As shown in Figure
\ref{offset}, the offsets for our SZ-selected sample are in agreement with
these results.  We find a strong anticorrelation between \csb\ and the BCG
offset as well as an absence of noncool cores with low offsets.  The two
parameters exhibit a Spearman's rank correlation coefficient of $\rho = -0.83$
with a $p$-value of $2.2 \times 10^{-4}$.  These results reinforce the model
that as clusters relax, BCGs settle in the center of the potential well and
cool cores become established.  Later, dynamical disturbances are capable of
removing a cool core.  Within this model, the clean anticorrelation reinforces
\csb\ as a parameter that accurately measures cluster cores at high redshifts.
The lack of ``relaxed,'' noncool-core clusters with small BCG offsets suggests
that neither preheating \citep[as proposed by][]{kaiser91,mccarthy08} nor
strong AGN feedback \citep[as reviewed in][]{fabian12,mcnamara12} are the most
dominant mechanisms of cool-core disruption in our sample.  Rather, this
anticorrelation points to recent mergers playing a major role by mixing or
shock-heating the ICM.

\section{Conclusions}\label{sec:conclusions}

We study the cool-core characteristics of an SZ-selected cluster sample
detected by the SPT survey.  The sample provides an alternative to
the existing X-ray-selected samples for studying the cool-core fraction at
high redshifts, which may be impacted by X-ray selection effects.

We evaluate the two parameters that are suitable for measuring cool cores at
high redshifts, the concentration of surface brightness (\csb) and the
cuspiness ($\alpha$), and examine their performance as a predictor of the
central cooling time $t_\mathrm{cool}$.  A Spearman's rank test shows
correlation coefficients between $\alpha$ and $t_\mathrm{cool}$ and between
\csb\ and $t_\mathrm{cool}$ to be $\rho_{\alpha} = -0.88$ and $\rho_{\csb} =
-0.84$, respectively.  However, \csb\ exhibits much smaller fractional
measurement errors.

Using \csb, we find evidence of two strong cool-core clusters in the SPT sample
at $z > 0.5$, among the first of their kind.  We rule out the hypothesis that
there are no such galaxy clusters at the 5.4$\sigma$ level.  With a sample of
nine high-redshift clusters, we show that the high-$z$ strong cool-core
fraction is greater than 7\% with 95\% confidence.  We compare the
distributions of the SPT sample and previous X-ray-selected samples with a K-S
test.  The result yields a $p$-value of 0.42, providing no statistically
significant evidence that the two samples are drawn from different
distributions.

We also evaluate the relationship between the strength of a cluster's cool core
and its dynamical state.  We find a strong anticorrelation between \csb\ and
the offset between the BCG and the X-ray centroid, which is related to the
dynamical state of the cluster.  While preheating or AGN feedback may be
responsible for the lack of cool cores in some clusters, this result suggests
that in our sample, the formation of cool cores is inhibited by merger events
in clusters, which cause turbulent mixing or shock-heating of the ICM.

The results of this study are based on a complete sample of SZ-selected
clusters with \chandra\ observations sampled from the first 178~deg$^2$ of the
SPT survey.  The statistics of our results will soon be dramatically improved
through a \chandra\ X-ray Visionary Project underway to observe the 80 most
significant clusters at $z > 0.4$ from the first 2000~deg$^2$ of the SPT
survey.

\acknowledgements

The South Pole Telescope program is supported by the National Science
Foundation through grant ANT-0638937.  The Munich group is supported by The
Cluster of Excellence ``Origin and Structure of the Universe'', funded by the
Excellence Initiative of the Federal Government of Germany, EXC project No.
153.  Galaxy cluster research at the University of Chicago is partially
supported by Chandra award No. GO2-13006A issued by the Chandra X-ray
Observatory Center.  Partial support is also provided by the NSF Physics
Frontier Center grant PHY-0114422 to the Kavli Institute of Cosmological
Physics at the University of Chicago, the Kavli Foundation, and the Gordon and
Betty Moore Foundation.  Galaxy cluster research at Harvard is supported by NSF
grant AST-1009012.  Galaxy cluster research at SAO is supported in part by NSF
grants AST-1009649 and MRI-0723073.  The McGill group acknowledges funding from
the National Sciences and Engineering Research Council of Canada, Canada
Research Chairs program, and the Canadian Institute for Advanced Research.
X-ray research at the CfA is supported through NASA contract NAS 8-03060.
Support for X-ray analysis is provided by the National Aeronautics and Space
Administration through Chandra award No. GO0-1114 issued by the Chandra X-ray
Observatory Center, which is operated by the Smithsonian Astrophysical
Observatory for and on behalf of the National Aeronautics Space Administration
under contract NAS8-03060.  R.J.F. is supported by a Clay Fellowship. B.A.B is
supported by a KICP Fellowship.  M.Bautz and M.M acknowledge support from
contract 2834-MIT-SAO-4018 from the Pennsylvania State University to the
Massachusetts Institute of Technology.  M.M acknowledges support from NASA
Hubble fellowship grant HST-HF-51308.01.  M.D. acknowledges support from an
Alfred P. Sloan Research Fellowship, C.J. acknowledges support from the
Smithsonian Institution, and B.S. acknowledges support from the Brinson
Foundation.  The authors thank Marcella Brusa for her helpful discussion.

\bibliography{../../BIBTEX/spt}
\bibliographystyle{apj}

\end{document}